\documentclass[preprint,showpacs,preprintnumbers,amsmath,amssymb]{revtex4}

\usepackage{graphicx}
\usepackage{dcolumn}
\usepackage{bm}

\begin{document}


\title{FRAGMENTATION CHANNELS OF RELATIVISTIC  $^7$Be NUCLEI IN PERIPHERAL INTERACTIONS}

\author{N.~G.~Peresadko}
   \affiliation{Lebedev Institute of Physics, Russian Academy of Sciences, Moscow, Russia} 
\author{V.~G.~Larionova}
   \affiliation{Lebedev Institute of Physics, Russian Academy of Sciences, Moscow, Russia} 
\author{Yu.~A.~Aleksandrov}
   \affiliation{Lebedev Institute of Physics, Russian Academy of Sciences, Moscow, Russia}
\author{S.~G.~Gerasimov}
   \affiliation{Lebedev Institute of Physics, Russian Academy of Sciences, Moscow, Russia} 
\author{V.~A.~Dronov}
   \affiliation{Lebedev Institute of Physics, Russian Academy of Sciences, Moscow, Russia} 
\author{S.~P.~Kharlamov}
   \affiliation{Lebedev Institute of Physics, Russian Academy of Sciences, Moscow, Russia} 
\author{V.~N.~Fetisov}
   \affiliation{Lebedev Institute of Physics, Russian Academy of Sciences, Moscow, Russia} 
\author{V.~Bradnova}
\affiliation{Joint Insitute for Nuclear Research, Dubna, Russia}
\author{S.~Vok\'al}
\affiliation{Joint Insitute for Nuclear Research, Dubna, Russia} 
\author{P.~I.~Zarubin} 
\homepage{http://becquerel.jinr.ru}
\homepage{http://pavel.jinr.ru}
\affiliation{Joint Insitute for Nuclear Research, Dubna, Russia} 
\author{I.~G.~Zarubina}
\affiliation{Joint Insitute for Nuclear Research, Dubna, Russia}
\author{A.~D.~Kovalenko}
\affiliation{Joint Insitute for Nuclear Research, Dubna, Russia}  
\author{A.~I.~Malakhov}
\affiliation{Joint Insitute for Nuclear Research, Dubna, Russia}    
\author{P.~A.~Rukoyatkin}
\affiliation{Joint Insitute for Nuclear Research, Dubna, Russia} 
\author{V.~V.~Rusakova}
\affiliation{Joint Insitute for Nuclear Research, Dubna, Russia}

\date{\today}

\begin{abstract}
\indent
Nuclei of $^7$Li were accelerated at the JINR Nuclotron. After the charge-exchange reaction 
involving these nuclei at an external target a second $^7$Be beam of energy 1.23A GeV was formed.
This beam was used to expose photo-emulsion chambers. The mean free path for inelastic $^7$Be interactions 
in emulsion $\lambda$=14.0$\pm$0.8 cm coincides within the errors with those for $^6$Li and $^7$Li nuclei. 
More than 10\% of the $^7$Be events are associated with the peripheral interactions in which the total charge 
of the relativistic fragments is equal to the charge of the $^7$Be and in which charged mesons are not produced.
An unusual ratio of the isotopes is revealed in the composition of the doubly charged $^7$Be fragments:
the number of $^3$He fragments is twice as large as that of $^4$He fragments. In 50\% of peripheral interactions, 
a $^7$Be nucleus  decays to two doubly charged fragments. The present paper gives  the channels of the $^7$Be
fragmentation to charged fragments. In 50\% of events, the $^7$Be fragmentation proceeds only to charged fragments 
involving no emission of neutrons. Of them, the $^3$He+$^4$He channel dominates, the $^4$He+d+p and $^6$Li+p
channels constitute 10\% each. Two events involving no emission of neutrons are registered  in the 3-body  $^3$He+t+p 
and $^3$He+d+d channels. The mean free path for the coherent dissociation of relativistic $^7$Be nuclei to 
$^3$He+$^4$He is 7$\pm$1 m. The particular features of the relativistic $^7$Be fragmentation  
in such peripheral interactions are explained by the $^3$He+$^4$He 2-cluster structure of the $^7$Be nucleus.
\par
\end{abstract}
 \pacs{21.45.+v,~23.60+e,~25.10.+s}

\maketitle
\section{\label{sec:level1}Exposure of emulsions  to a $^7$Be beam}
\indent In order to form a $^7$Be beam at the JINR Nuclotron, $^7$Li nuclei were accelerated to an energy of 2.87Z GeV.
The extracted $^7$Li beam was delivered to a target of organic glass. The $^7$Be nuclei produced at the target due to
a $^7$Li charge exchange reaction were focused by magnetic elements and a secondary beam was thus formed. The charges of 
the particles of this beam were determined by particle energy losses  in a scintillation monitor. According to these 
measurements, the admixture of Z=3 particles in the beam was 7\% of the number of Z=4 particles.
\par
\indent Photo-emulsion chambers assembled out of emulsion layers, the thickness and dimension of which were 550 $\mu$m and 
10$\times$20 cm$^2$, respectively, were exposed to the $^7$Be beam \cite{web}. The exposed emulsion layers were parallel
to a beam axis. The standard BR-2 emulsion  was used in which singly and doubly charged relativistic particles were well
identified  by sight. The particle charges larger than 2 were determined by the density of the gaps in the particle tracks
using a computer analysis of a digitized image of a vision field on a microscope with an automatic track lock-on.
According to the measurements of the particle charges in emulsion, the triply charged particles in the beam are about 15\%
of the particles with Z=4. $^7$Be interactions in emulsion were sought with the aid of microscopes by selecting the 
particle tracks  with the largest ionization density.          
\par

\section{\label{sec:level2}The mean free path for inelastic $^7$Be interactions in emulsion}

\indent Inelastic nucleus-nucleus interactions were sought in emulsion by viewing the tracks of particles starting with
their entrance into the emulsion by means of a microscope with a magnification of $\times$900. In order to determine the 
mean free path of $^7$Be inelastic interactions in emulsion $\lambda$($^7$Be), use was made  of 294 inelastic interactions 
detected  over the viewed track length  of 41.222 m in one emulsion chamber. Table \ref{tab:1} gives the result obtained for the 
$^7$Be nucleus and the values of the mean free path of inelastic interactions of $^6$Li and $^7$Li in emulsion determined
in \cite{El-Sharkawy,Lepekhin,Adamovich99,Adamovich04,Nady}. The measured values for  all these nuclei are within the errors nearly identical. The Table gives also the 
values calculated by a geometric model with a set of parameters employed for the description of the mean free
path of the inelastic interactions of nuclei  of homogeneous density in emulsion. The fact that the experimental values of 
the mean free path for all these nuclei are smaller than the calculated ones is accounted for by an additional contribution 
from peripheral inelastic interactions of nuclei having a loosely bound cluster structure.
\par

\begin{table}
\caption{\label{tab:1}Mean free paths of $^6$Li, $^7$Li, and $^7$Be nuclei for inelastic interactions in emulsion.}

\begin{tabular}{c|c|c|c|c}
\hline\noalign{\smallskip}
\hline\noalign{\smallskip}
	~~Nucleus	~~      & ~~$\lambda_{exp}$, cm~~& ~~$\lambda_{calc}$, cm~~ & ~~Energy~~ & ~~Paper ~~ \\
	~~$^6$Li~~	    & ~~14.1$\pm$0.4~~	& ~~16.5-17.2~~  & ~~27~~         & ~~\cite{Adamovich99}~~ \\
	~~$^7$Li~~  & ~~14.3$\pm$0.4~~    & ~~16.0-16.3~~  & ~~21~~         & ~~\cite{Adamovich04}~~ \\
	~~$^7$Be~~	    & ~~14.8$\pm$0.8~~	    & ~~16.0-16.3~~	  & ~~8þ6~~	       & ~~\cite{Adamovich04}~~ \\

\hline\noalign{\smallskip}
\hline\noalign{\smallskip}
\end{tabular}
\end{table}

\section{\label{sec:level3}The isotopic composition of fragments and the $^7$Be fragmentation channels 
	in peripheral interactions of $^7$Be nuclei in emulsion}
\indent Of 1400 inelastic nucleus-nucleus interactions detected, more than 200 are peripheral interactions in which the 
total charge Q of relativistic particles  with 15$^\circ$ emission angle is equal to the charge of the primary $^7$Be nucleus. 
In about 150 peripheral interactions there is observed no charged meson production. In such interactions, the particular
features of the structure of the nucleus most strongly affect the character of the nuclear fragmentation and, first of all, 
the charge and mass composition of the fragments. Table \ref{tab:2} displays the charge topology of such events. The events involving
no target fragments (n$_b$=0) are separated from the events  involving one or a few  fragments (n$_b>$0). In a half of the 
interactions each of them contains two doubly charged fragments, while in the other half each event contains one helium and
two singly-charged fragments. 10\% of events contain a relativistic Li nucleus accompanied by a single-charged fragment.
A large fraction of events which are due to the dissociation of $^7$Be nuclei into two helium fragments suggests that a
clustering of this type in the $^7$Be structure is very possible.     
\par
\indent The isotopic composition of fragments was studied by measuring the multiple Coulomb scattering of particles in
emulsion. The values of p$\beta$c, where p is the momentum and  $\beta$ the particle velocity, were determined. The momenta 
of singly and doubly charged particles were measured in 240 interactions of $^7$Be nuclei with the emulsion nuclei. The 
experimental p$\beta$c distribution of relativistic doubly charged particles is satisfactorily approximated by two Gauss 
functions with peaks at p$\beta$c equal to 4.5 GeV and 6.3 GeV. A relative fraction of $^3$He and $^4$He fragments estimated
over the areas covered by the approximating curves  is 70\% and 30\%, respectively. In the interactions of all other 
relativistic nuclei in emulsion which were investigated earlier the fraction of $^4$He fragments is larger than that of 
$^3$He fragments. Such a anomalous ratio of the He isotopes observed in $^7$Be interactions is explained by the 2-cluster 
structure of the $^7$Be nucleus in which the nucleons not involved to the $\alpha$ particle core are bound into a $^3$He 
cluster. The p$\beta$c distribution of singly charged relativistic particles in an interval to p$\beta$c=5 GeV is 
satisfactorily described by two Gauss functions with peaks at p$\beta$c equal to 1.5 GeV and 3.2 Gev. The proton to
deuteron ratio is estimated to be 3:1. The number of particles of momenta higher than 5 GeV/c constitutes about 2\%
 of the total number of singly charged fragments. The results of these measurements were used to determine the fragment 
mass in each event and identify the $^7$Be fragmentation channels.
\par

\begin{table}
\caption{\label{tab:2}Fragment charge composition in events Q=4.}

\begin{tabular}{c|c|c}
\hline\noalign{\smallskip}
\hline\noalign{\smallskip}
	~~Relativistic fragments 	~~      & ~~Target ~~& ~~Number of events~~ \\
	~~2He~~	    & ~~n$_{b}=$0~~	& ~~41~~  \\
	~~2He~~	    & ~~n$_{b}>$0~~	& ~~18~~  \\
	~~He+2H~~	    & ~~n$_{b}=$0~~	& ~~42~~  \\
	~~He+2H~~	    & ~~n$_{b}>$0~~	& ~~33~~  \\
	~~4H~~	    & ~~n$_{b}=$0~~	& ~~2~~  \\
	~~4H~~	    & ~~n$_{b}=$1~~	& ~~1~~  \\
	~~Li+H~~	    & ~~n$_{b}=$0~~	& ~~9~~  \\
	~~Li+H~~	    & ~~n$_{b}>$1~~	& ~~3~~  \\
	~~Total~~	    & ~~ ~~	& ~~149~~  \\
\hline\noalign{\smallskip}
\hline\noalign{\smallskip}
\end{tabular}
\end{table}

\indent Table \ref{tab:3} presents the numbers of the events detected in various channels of the $^7$Be fragmentation. Of them, the
$^3$He+$^4$He channel noticeably dominates, the channels $^4$He+d+p and $^6$Li+p constitute 10\% each. Two events involving
no emission of neutrons in the three-body channels $^3$He+t+p and $^3$He+d+d were registered. The reaction of 
charge-exchange of $^7$Be nuclei to $^7$Li nuclei was not detected among the events not accompanied by other secondary 
charged particles.
\par
\begin{table}
\caption{\label{tab:3}$^7$Be fragmentation channel (number of events)}

\begin{tabular}{c|c|c|c|c|c|c|c|c|c}
\hline\noalign{\smallskip}
\hline\noalign{\smallskip}
Channel&2He&2He&He+2H&He+2H&4H&4H&Li+H&Li+H&Sum\\

~&n$_{b}=$0&n$_{b}>$0&n$_{b}=$0&n$_{b}>$0&n$_{b}=$0&n$_{b}>$0&n$_{b}=$0&n$_{b}>$0&~\\
~$^3$He+$^4$He&30&11&  &  &  &  &  &  & 41\\
~$^3$He+$^3$He&11& 7&  &  &  &  &  &  & 18\\
~$^4$He+2p    &  &  &13& 9&  &  &  &  & 22\\
$^4$He+d+p   &  &  &10& 5&  &  &  &  & 15\\
~$^3$He+2p    &  &  & 9& 9&  &  &  &  & 18\\
$^3$He+d+p   &  &  & 8&10&  &  &  &  & 18\\
~$^3$He+2d    &  &  & 1&  &  &  &  &  &  1\\
$^3$He+t+p   &  &  & 1&  &  &  &  &  &  1\\ 
~3p+d         &  &  &  &  & 2&  &  &  &  2\\
~2p+2d        &  &  &  &  & 1&  &  &  &  1\\
~$^6$Li+p     &  &  &  &  &  &  & 9& 3& 12\\
Sum          &41&18&42&33& 2& 1& 9& 3&149\\

\hline\noalign{\smallskip}
\hline\noalign{\smallskip}
\end{tabular}

\end{table}
\indent The events containing only two helium fragments are given in Fig.1 by points  the coordinates of which are the 
measured p$\beta$c fragment values. The larger value of p$\beta$c (max.) is taken as an abscissa and the smaller value of
p$\beta$c (min.) – as an ordinate. All the events over the ordinate axis are located below 5 GeV. This value implies the
lower boundary of p$\beta$c for $^4$He nuclei. The $^3$He+$^3$He events are seen to be to the left of the p$\beta$c(max.) boundary,
while  the $^3$He+$^4$He events are to the right of the boundary. The fraction of the $^3$He+$^4$He channel of all the 
$^7$Be dissociation events, which amounts to 30\%, may be sought of as an estimate of the lower value of the probability of
this configuration in the $^7$Be nucleus. The mean free path for the coherent dissociation of relativistic $^7$Be nuclei to 
$^3$He+$^4$He in emulsion is 7$\pm$1 m . The mean free paths of $^6$Li, $^7$Li and $^7$Be for the 2-particle channels of
the coherent dissociation involving no neutron emission  have close values. Direct estimates of the fact that the $^6$Li
nucleus can be in the state of an $\alpha$ particle core  and a quasi-free deuteron cluster, which were obtained by a 
1 GeV $\pi^-$ mesons sounding of a $^7$Li target in experiment \cite{Abramov}, exceed 0.75. 
\par

\begin{figure}

\includegraphics[width=100mm]{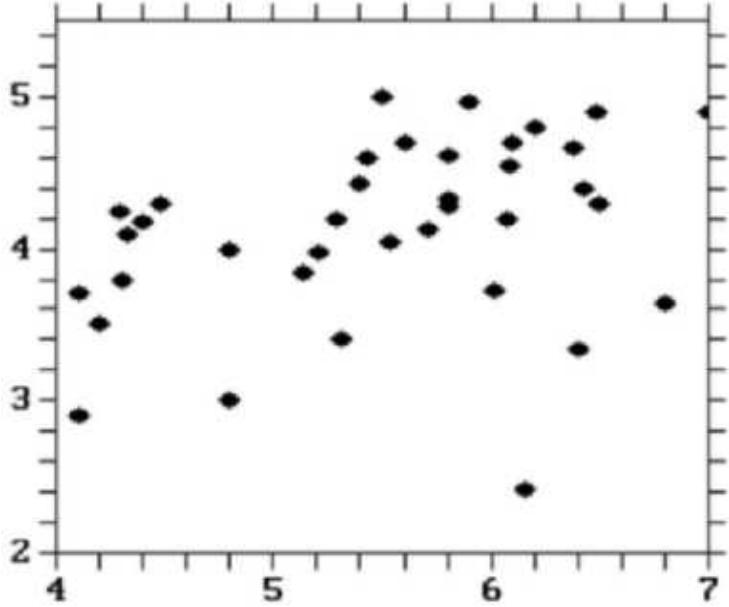}
\caption{\label{fig:1} Distribution of 2He events over the fragment momenta. The events are indicated as dots with 
coordinates corresponding to measured p$\beta$c values. The maximum p$\beta$c value is attributed to the ordinate,
 and the minimum one to the abscissa.}

\end{figure}
\indent Fig. 2 shows the distribution of the 
$^3$He+$^4$He channel events depending on E= 1.59+E$_t$, where E$_t$ is equal to the transverse kinetic energy of the 
fragments and the value 1.59 MeV is equal to the threshold energy of the channel. In more than 80\% of events the E values 
do not exceed  10 MeV. The same energy region is also occupied by the $^7$Be excitation levels the positions of which are
indicated by arrows. The separation of individual levels in the experimental distribution is not observed. 
\par
\indent The fragment 
system energy may also be characterized by the transverse momenta of fragments in the coordinate frame of reference 
associated with a fragmenting nucleus. The difference in the average values of the fragment momenta  for mirror nuclei may
 be viewed as a manifestation  of  the influence of the Coulomb interaction of charged clusters in nuclei and in the process
 of fragmentation of these nuclei. The mean values of the transverse momenta of the fragments in the $^3$He+$^4$He channel 
in their c.m.s. is 147$\pm$5 MeV/c. A noticeable exceeding of this value with respect to the average values of the 
transverse momenta of the fragments in the $^7$Li$\rightarrow^3$H+$^4$He  fragmentation channel  equal to 108 MeV/c may be 
treated as an effect of the Coulomb interaction of the clusters in these nuclei. 
\par

\begin{figure}
\includegraphics[width=100mm]{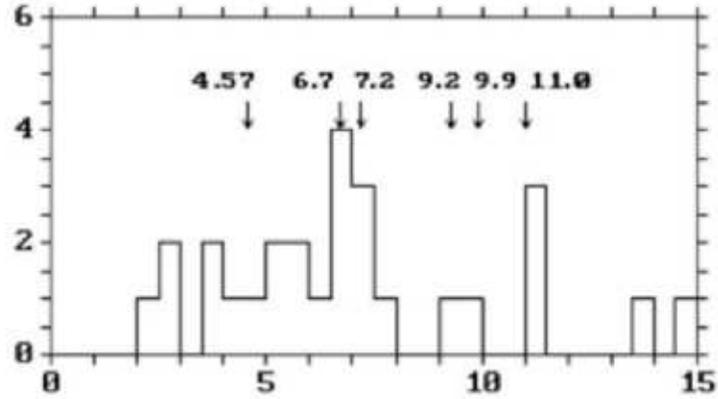} 
\caption{\label{fig:2}Distribution of $^3$He+$^4$He events over the value E.}

\end{figure}
\indent Fig.3 shows the distribution of the $\psi$
angles between the $^3$He and $^3$He fragments in the asymuthal plane in $^3$He+$^4$H events. Large angles between the 
fragments are dominant. This distribution is, to a large extent, defined by the momenta transferred to fragmenting nuclei. 
The $\psi$ angles close to 180$^\circ$ correlate with small momenta transferred to the $^7$Be nucleus. A relatively large
number of the events close to 180$^\circ$ angles and having low values of the transferred momenta may be due to the
contribution of the Coulomb dissociation of $^7$Be by heavy nuclei of the emulsion.
\par

\begin{figure}
\includegraphics[width=100mm]{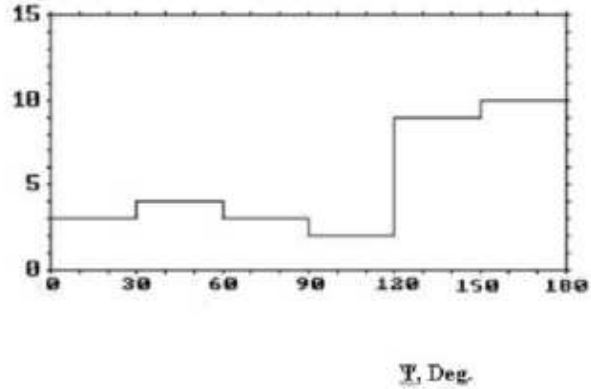}
\caption{\label{fig:3} Distribution of azimuthal angles $\psi$ between fragments $^3$He and $^4$He.}

\end{figure}
\section{\label{sec:level4}Conclusions}
\indent 
Thee secondary $^7$Be beam of energy 1.23A GeV was formed at the JINR Nuclotron  by means of the charge exchange reaction 
involving $^7$Li nuclei in an external target. This beam was used to expose nuclear emulsion stacks. The mean free path for
inelastic $^7$Be interactions in emulsion $\lambda$($^7$Be)=14.0$\pm$0.8 coincides within the errors with those for
inelastic $^6$Li and $^7$Li interactions. Of 1400 inelastic $^7$Be interactions with emulsion nuclei, there were detected
149 peripheral interactions in which the total charge of the relativistic $^7$Be fragments was equal to the $^7$Be charge 
and in which charged mesons were not produced. In 50\% of such interactions, each event involves two helium fragments. 
An unusual ratio of isotopes is observed in the composition of the $^7$Be doubly charged fragments: the fraction of $^3$He 
fragments is as twice as large than that of $^4$He fragments. The channels of fragmentation are given in the present
paper. In 50\% of the events, the fragmentation proceeds only to charged framents involving no neutron emission. Of them,
the $^3$He+$^4$H channel is dominant. The mean free path for the coherent dissociation of relativistic $^7$Be nuclei to 
$^3$He+$^4$He in emulsion is 7$\pm$1 m. A $^7$Be dissociation process is registered in the $^6$Li+p channel. 
The main characteristics of the relativistic $^7$Be fragmentation are determined by the 2-helium cluster structure of the
$^7$Be nucleus. The mean value of the fragment transverse moments 147$\pm$5 MeV/c in the channel of the coherent 
fragmentation of $^7$Be nuclei to $^3$He+$^4$He is observed to exceed the value 108$\pm$2 MeV/c for the channel of $^7$Li 
dissociation to $^3$H+$^4$He. This fact may be associated with the Coulomb interaction of fragments in these 
processes. A relatively large number of events having   $\psi$ angles in the region of 180$^\circ$ and low transverse 
momenta may be due to the contribution of the Coulomb dissociation of $^7$Be nuclei on heavy emulsion nuclei.

\begin{acknowledgments}
 \indent  The authors thank the JINR Nuclotron staff for the emulsion exposure and to the LHE JINR chemist
 group for an emulsion proceeding. They are thankful to A. B. Antipova, A. V. Pisetskaya, and L. N. Shesterkina of Lebedev
 Institute for scanning and measuremnets of nucleus-nucleus interactions. The authors are indebted to Profs. F. G. Lepekhin
 and M. M. Chernyavsky for discussions of the work.  
\par 
\indent The work was supported by the Russian Foundation for Basic Research,
 grant 96-159623, 02-02-164-12a.\par
 
\end{acknowledgments}

\end{document}